\documentclass[reprint, superscriptaddress, amsmath, amssymb, aps,showpacs]{revtex4-1}

\usepackage{graphicx}
\usepackage{dcolumn}
\usepackage{bm}
\usepackage{mathtools}

\begin{document}

\title{Visualization of superposition states and Raman processes\\ with two-dimensional atomic deflection}

\author{Gor A. Abovyan}
\email{gor.abovyan@ysu.am}
\affiliation{Yerevan State University, A. Manookyan 1, 0025, Yerevan, Armenia}
\affiliation{Institute for Physical Research, Ashtarak-2, 0203, Ashtarak, Armenia}

\author{Gagik P. Djotyan}
\email{djotjan@rmki.kfki.hu}
\affiliation{Institute for Particle and Nuclear Physics of the Hungarian Academy of Sciences, Konkoly-Thege  Mikl\'{o}s \'{u}t 29-33, H-1121 Budapest, Hungary}

\author{Gagik Yu. Kryuchkyan}
\email{kryuchkyan@ysu.am}
\affiliation{Yerevan State University, A. Manookyan 1, 0025, Yerevan, Armenia}
\affiliation{Institute for Physical Research, Ashtarak-2, 0203, Ashtarak, Armenia}



\date{\today}

\begin{abstract}

Deflection of atoms in $\Lambda$-type configuration passing through two crossed standing light waves is proposed for probing and visualization of atomic superposition states. For this goal, we use both the large-dispersive and Raman-resonant regimes of atom-field interaction giving rise to a position-dependent phase shifts of fields and perform double simultaneous spatial measurements on an atom. In this way, it is demonstrated that the deflection spatial patterns of atoms in $\Lambda$-configuration passing through modes of standing waves are essentially modified if the atoms are initially prepared in a coherent superposition of its low levels states as well as when the superposition states are created during the process of deflection. The similar results take place for the joint momentum distribution of atoms. Further, considering both one-photon and two-photon excitation regimes of $\Lambda$-atoms we also illustrate that the two-dimensional patterns of defected atoms qualitatively reflects the efficiency of the Raman processes. 
\end{abstract}

\pacs{37.30.+i, 42.50.Dv, 42.50.Pq, 42.50.St}
\maketitle


\section{\label{Introduction}Introduction}

The ability to prepare atomic systems in superposition states is important both in fundamental studies of quantum mechanics as well as for various technological applications including the field of quantum information and quantum lithography. It is evident, that application of strong, near-resonant to atomic transitions laser light may result in the production and probing of coherent superposition of atomic or molecular states. In this way, many experiments have been proposed or realized, particularly with single trapped ion \cite{Monr,*Myatt2000}, or microwave cavity quantum electrodynamics \cite{Raim}, with a single Rydberg atom coupled to a single field mode.  Creation of such quantum states has also been realized for molecular systems \cite{Arndt} including large organic molecules \cite{Gerlich}, for atomic ensembles \cite{sagafsg} and even virus \cite{Romero}. Preparation of the atoms or molecules in the coherent superposition states may lead to substantial changes in optical properties of a medium composed of the particles. Some of the most spectacular examples are: electromagnetically induced transparency (EIT) with extreme change in the group velocity of laser pulses, even including complete stopping of laser pulses (see, review papers \cite{Lukin,Fleisch}), enhancement of the efficiency of nonlinear optical processes \cite{Jain,Lukin1998} and  writing and storage of optical information in meta-stable quantum states \cite{Djotyan1,Djotyan2,Djotyan}. The preparation of quantum coherence has also become of paramount importance for the growing field of quantum information science \cite{Nielsen,Kok,Schumacher}. There exist some techniques for probing the quantum interference based on interaction of an atom in a superposition state with field modes. In this way, a final detection is realized by homodyne measurement of states of light field after its interaction with atomic system as well as by the methods of quantum tomography \cite{Kienle}. Particularly, much work has been focused on applications and developments of the technique of quantum tomography for atomic beams, (see \cite{Frey} and Refs. therein).

One of the basic processes of atom optics is the deflection of atomic beams when interacting with a standing light wave inside an optical cavity. In this paper we demonstrate that production of atomic superposition states is qualitatively displayed in two-dimension patterns of deflected atoms on a two crossed standing waves. The analysis is done in quantum treatment for three-level atoms in $\Lambda$ - configuration interacting with two crossed standing light waves. We demonstrate that the deflection patterns of the atomic beam passing through two crossed standing light waves are modified if the atoms are initially prepared in coherent superposition of lower levels states or as well as when the superposition states are created during the process of deflection. Our other goal is to understand how Raman processes under the general two-photon resonance condition are exhibited in the atomic deflection patterns. Considering different interaction regimes of $\Lambda$ - system with off-resonant standing waves we have demonstrated that the deflection patterns, in the transverse plane to the direction of the center of atomic mass motion, are essentially different for the case of one-photon and two-photon Raman interactions. 

The approach proposed relies on the problem of atomic spatial localization. In general, the precise spatial measurement and localization of quantum particles has been a subject of considerable interest since the discussion on Heisenberg's famous microscope. A particular class of quantum optical localization schemes suitable to determine the position of a quantum particle on a subwavelength scale makes use of standing wave driving fields (see,\cite{Storey1992,*Storey1993,Kunze,Kien,Hero,Pasp,Sahrai,*Kapale,AgaKap}).

 It is well known that when atoms pass a standing-wave cavity mode the strength of interaction with field depends on the positions of the atoms. Thus, quadrature phase measurement on the field leads to strong localization of the atomic position below a wavelength of the field in the cavity \cite{Storey1992,*Storey1993}. Recently, atomic localization and centre-of-mass wave function measurement via multiple simultaneous dispersive interactions of atoms with different standing-wave fields have been investigated \cite{Evers} in addition to the well-known results for a single-mode cavity \cite{Storey1992,Storey1993,Kunze,Kien,Hero,Pasp,Sahrai,*Kapale,AgaKap}. Note, that analogous scheme of atomic deflection has been recently considered for investigation of spatial entanglement in deflection of $V$-type atoms \cite{Kryuchkyan,GXLi} as well as $\Lambda$-type atoms \cite{Gor}. Thus, we apply the measurement-induced localization procedure for our scheme, calculating the conditional position distribution of atoms while considering two-mode field to be in a given reference quadrature-phase state. As it will be shown here, for narrow initial position distributions of atoms our scheme permits producing controllable pattern structures with feature spacing smaller than a wavelength of the light in the cavity. 

For completeness, we also discuss the visualization of atomic superposition states in the momentum space. In this direction the distribution of deflected atoms in terms of the transverse atomic moment is calculated. It is evident that for narrow initial atomic wave packets wider distributions in the momentum space will be realized.

This paper is organized as follows. In section \ref{sec:2} we introduce the system and obtain the formulas for the conditional position distribution of atoms. In section \ref{sec:3} we present the results for probing of superposition states for both scheme of measurement, for near to the cavity zones and  outside the cavity in far-diffraction zones. The momentum distributions are also calculated.  Section \ref{Raman} is devoted to visualization of the Raman resonance.  Section \ref{Sum} concludes the paper.

\section{\label{sec:2}Atomic deflection in the presence of two-photon, Raman processes}

Let us consider the quantum dynamics of a three-level atom with a $\Lambda$-type configuration of energy levels moving along the $z$-direction and passing through cavities that involve two crossed one-mode standing waves (see, Fig.1). Atomic beam is adjusted so that only one atom interacts with the cavity electromagnetic field at a time and position patterns of deflected atoms in the $x-y$ plane are measured. The transition between the two lower levels $|1\rangle$ and $|2\rangle$ is dipole forbidden and the transition from the upper level $|3\rangle$ to any of the lower $|1\rangle$ and $|2\rangle$ levels is allowed. We focus more specifically on the dispersive limit where the detuning between the two standing wave frequencies and the corresponding atomic transition frequencies are large compared with the Rabi frequencies. We also neglect the atomic damping during the time an atom interacts with the fields.  

\begin{figure}[h]
 \begin{math}
 \begin{array}{cc}
    \includegraphics[height=3.3cm]{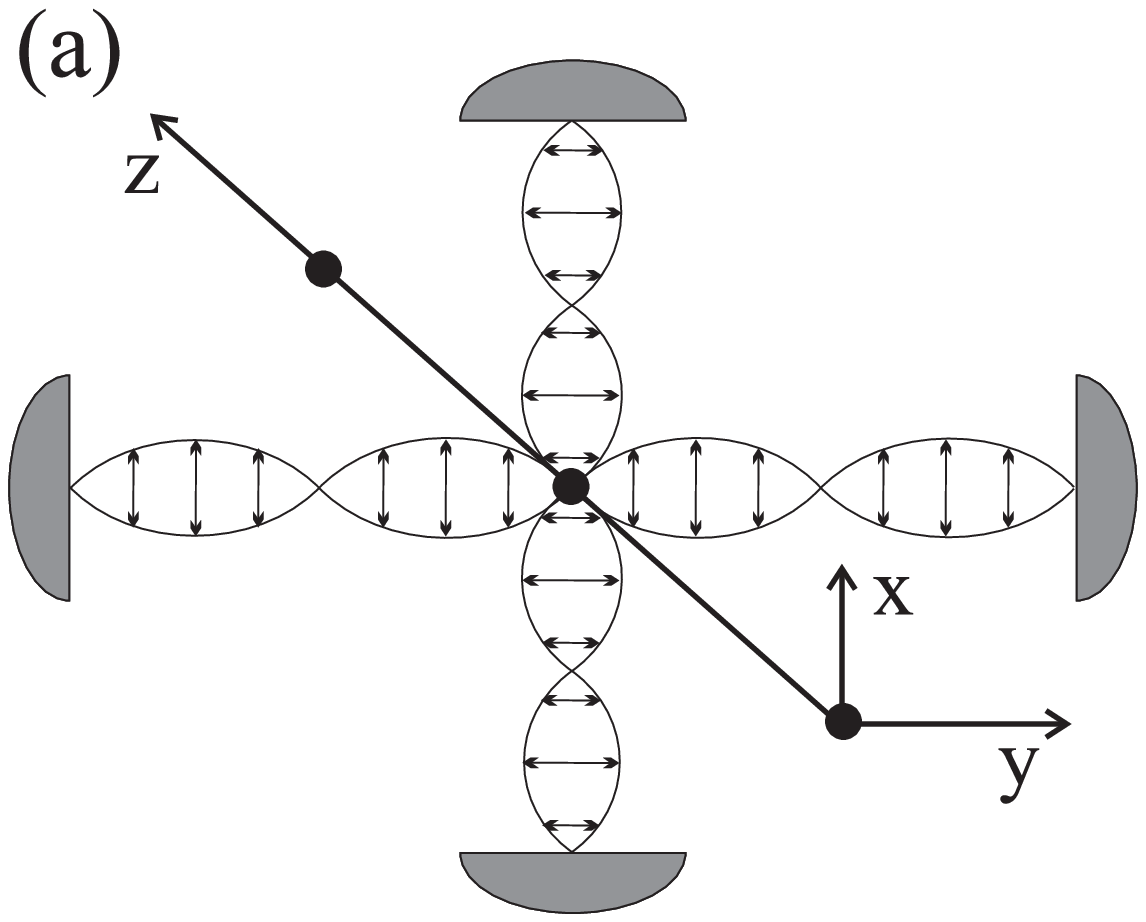} & 
    \includegraphics[height=3.2cm]{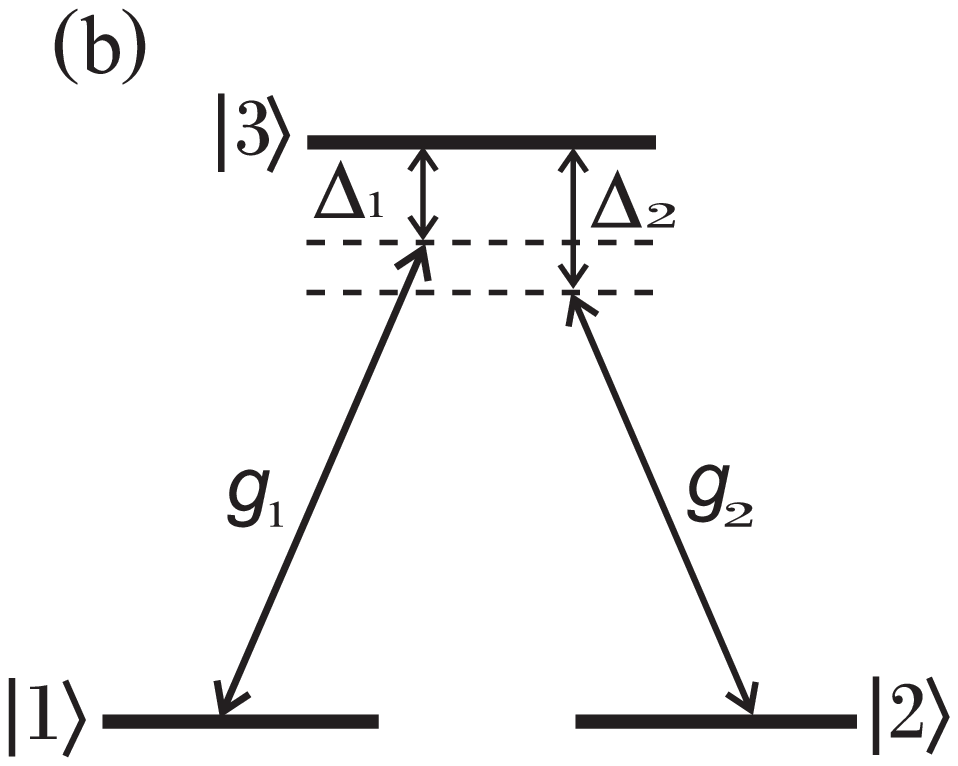}\\
 \end{array}
 \end{math}
  \caption{\label{atom} Schematic diagrams showing the investigated model. (a) The atomic beam crossing the interaction region. (b) Energetic levels of a $\Lambda$-type of atoms with equal energies of sublevels interacting with modes of opposite circular polarizations with coupling constants $g_{1}$ and $g_{2}$.}
\end{figure}

We consider two standard types of Hamiltonian depending on two detunings $\Delta_{1}$ and $\Delta_{2}$. If the frequencies of modes and the duration $\tau$ of atom-field interaction are adjusted so that $\Delta_{1}-\Delta_{2}\ll\pi/\tau$, the case of two-photon resonance $\Delta_{1}=\Delta_{2}=\Delta$ can be realized with the following effective Hamiltonian 
\begin{eqnarray}\label{RamHam}
H_{eff}=\sum_{i=1,2}\frac{\hbar g_{i}^{2}}{\Delta}a_{i}^{\dag}a_{i}\sigma_{ii} & \nonumber\\
+ \frac{\hbar g_{1}g_{2}}{\Delta} &  \left(a_{1}^{\dag}a_{2}\sigma_{12}+a_{1}a_{2}^{\dag}\sigma_{21}\right).
\end{eqnarray}
Here $a_{i}$ and $a_{i}^{\dag}$ are the annihilation and creation operators of the $i$th mode, while $\sigma_{ij}=|i\rangle\langle j|$ is the corresponding transition operator of $\Lambda$-atom. The couplings of the atom to the two modes are determined by the spatial mode functions $g_{1}(x)=g_{01}\sin(k_{1}x)$ and $g_{2}(y)=g_{02}\sin(k_{2}y)$, where $g_{0i}=E_{0}\vec{e}_{i}\langle i|\vec{d}|3\rangle$, $i=1,2$ and $k_{i}$ is the wave vector of $i$th mode. $E_{0}$ is the so-called electric field per photon. $\vec{e}_{1}$ and $\vec{e}_{2}$ are the polarization vectors, while $\langle 1|\vec{d}|3\rangle$ and $\langle 2|\vec{d}|3\rangle$ are the dipole moments of the corresponding atomic transitions. The last term in the expression (\ref{RamHam}) describes the connection between the two interaction channels due to two-photon (Raman) transitions between $|1\rangle$ and $|2\rangle$ levels. When the condition of the Raman resonance is not carried out and contribution of the Raman transitions may be neglected, the interaction Hamiltonian reads as
\begin{equation}\label{OutRaman}
H_{eff}=\sum_{i=1,2}\frac{\hbar g_{i}^{2}}{\Delta_{i}}a_{i}^{\dag}a_{i}\sigma_{ii}.
\end{equation}
The initial position distribution of atoms at the entrance of cavities is assumed to be Gaussian, i.e.
\begin{equation}
|f(x,y)|^{2}=\frac{1}{2\pi\Delta x\Delta y}e^{-\frac{(x-\langle x\rangle)^{2}}{2(\Delta x)^{2}}}e^{-\frac{(y-\langle y\rangle)^{2}}{2(\Delta y)^{2}}}
\end{equation}
with the widths $\Delta x=\sqrt{\langle (x-\langle x\rangle)^{2}\rangle}$ and $\Delta y=\sqrt{\langle (y-\langle y\rangle)^{2}\rangle}$ centered at the nodes of both waves, while the initial atomic states is considered in general as $a|1\rangle+b|2\rangle$ with $a$ and $b$ being the weights of the atomic lower states in the coherent superposition. The cavity modes are assumed to be initially in two-mode coherent state
\begin{equation}
|field\rangle=|\alpha_{1}\rangle_{1}|\alpha_{2}\rangle_{2},\text{     }|\alpha\rangle_{i}=e^{-N/2}\sum_{n}\frac{\alpha^{n}}{\sqrt{n!}}|n\rangle_{i},
\end{equation}
where $|n\rangle_{i}$ are the Fock states for $i$th mode and $N=|\alpha|^{2}$. In this case, the state vector of this system at time $t$ will have the form 
\begin{equation}\label{FinalS}
|\Psi(t)\rangle=\int d x d y \sum_{i=1,2}\sum_{n,m=0}^{\infty}\Phi_{n,m}^{(i)}(x,y,t)
|n\rangle_{1}|m\rangle_{2}|i\rangle|x,y\rangle, 
\end{equation}
where the amplitudes $\Phi_{n,m}^{(i)}(x,y,t)$ can be derived from the following equation and its corresponding initial conditions
\begin{equation}\label{Shred}
\left\{
\begin{split}
\mathbf{i}\hbar\frac{\partial}{\partial t}|\Psi(t)\rangle & =\hat{H}_{eff}|\Psi(t)\rangle,\\
|\Psi(t=0)\rangle & =\int dxdy f(x,y)|x,y\rangle\otimes(a|1\rangle+b|2\rangle)\\
& \otimes|\alpha_{1}\rangle_{1}|\alpha_{2}\rangle_{2}
\end{split}
\right.
\end{equation}
by substituting $|\Psi(t)\rangle$ in (\ref{Shred}) with the expression (\ref{FinalS}).
In the case of two-photon resonance the amplitudes are written as
\begin{subequations}
\label{Phis}
\begin{eqnarray}
& \Phi_{n,m}^{(1)}(x,y,t)=f(x,y)\nonumber\\
& \times\left\{
a\cdot C_{n,m}+G_{1}\cdot\left(\exp\left[-\frac{\mathbf{i}t}{\Delta}\Omega_{n,m+1}\right]-1\right)
\right\},\\
& \Phi_{n,m}^{(2)}(x,y,t)=f(x,y)\nonumber\\
& \times\left\{
b\cdot C_{n,m}+G_{2}\cdot\left(\exp\left[-\frac{\mathbf{i}t}{\Delta}\Omega_{n+1,m}\right]-1\right)
\right\},
\end{eqnarray}
\end{subequations}
where
\begin{subequations}
\label{ABs}
\begin{eqnarray}
G_{1}=\frac{g_{1}g_{2}\sqrt{n(m+1)}}{\Omega_{n,m+1}}C_{n-1,m+1}\cdot b\nonumber\\
+\frac{g_{1}^{2}n}{\Omega_{n,m+1}}C_{n,m}\cdot a,\\
G_{2}=\frac{g_{1}g_{2}\sqrt{(n+1)m}}{\Omega_{n+1,m}}C_{n+1,m-1}\cdot a\nonumber\\
+\frac{g_{2}^{2}m}{\Omega_{n+1,m}}C_{n,m}\cdot b,
\end{eqnarray}
\end{subequations}
$\Omega_{n,m}=g_{1}^{2}n+g_{2}^{2}m$ is the position-dependent Rabi frequency and 
\begin{equation}
C_{n,m}=e^{-(|\alpha_{1}|^{2}+|\alpha_{2}|^{2})/2}\frac{\alpha_{1}^{n}\alpha_{2}^{m}}{\sqrt{n!m!}}.
\end{equation}
The functions $\Phi_{n,m}^{(i)}(x,y,t)$ describe amplitude distribution for the position of atom in the $(x,y)$ plane. They are proportional to the atomic initial position distribution and also display not trivial spatial features due to position dependent phase shifts acquired by $\Lambda$ - atom passing through the standing waves. 

It is well known \cite{Frey,Storey1992} that the measurement of the phase shift of the cavity fields can be interpreted as a quantum spatial localization of the atom. In this way, below we investigate deflection of atoms with simultaneous quadrature measurement of the field. We have studied this problem in details for one-dimensional atomic deflection as well as for two-dimensional case in the dispersive limit assuming negligibly small the contribution of the Raman transitions in the system. In this case, there is no exchange of energy between the field and the atom, i.e. the interaction does not change the internal atomic state. This situation is cardinally changed for the $\Lambda$ - atom under two-photon resonance condition. Indeed, in this case both amplitudes $\Phi_{n,m}^{(1)}(x,y,t)$ and $\Phi_{n,m}^{(2)}(x,y,t)$ corresponding to two atomic states $|1\rangle$ and $|2\rangle$ govern the spatial distribution of deflected $\Lambda$ - atoms. Thus, the interaction between atom and intracavity field beside the initial position of the atom depends also on its internal states. The latter allows visualization of the atomic coherence in final position distribution of the atoms.

\section{\label{sec:3}Probing and visualization of the atomic coherence by the deflection patterns}

\subsection{Position distributions}

\begin{figure*}[!ht]
\includegraphics[width=17.5cm]{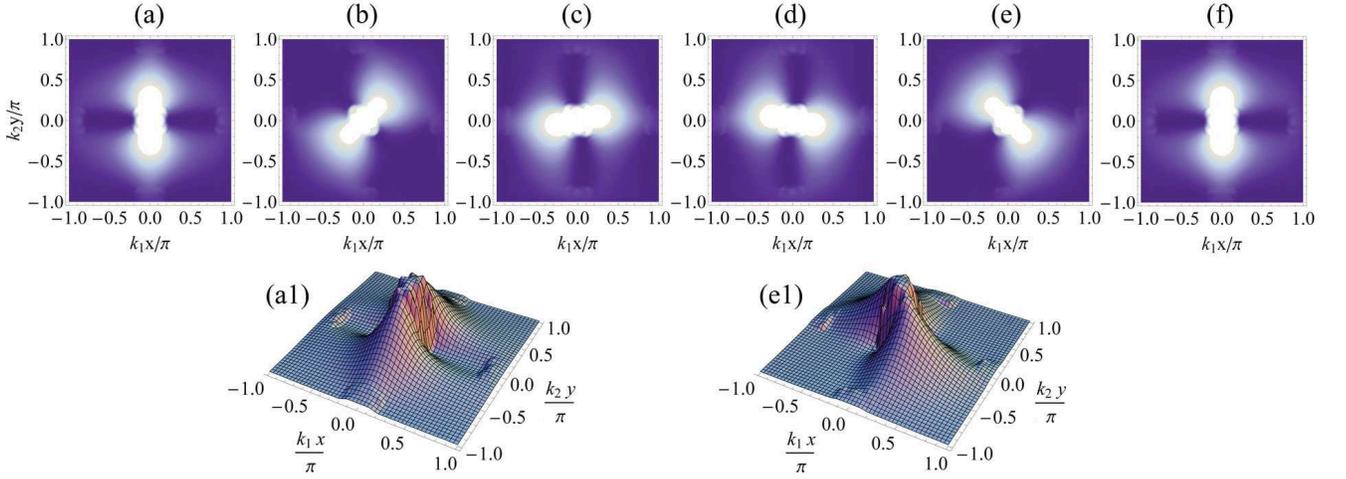}
\caption{\label{rotation}A set of diagrams showing the circular dependence of atomic distribution on initial internal state. Bright colors indicate high probability, dark colors low probability. The parameters are $\theta_{1}=\theta_{2}=0$, $\alpha_{1}=\alpha_{2}=2$, $\chi_{\theta_{1}}=\chi_{\theta_{2}}=4$ and $\Delta x=\Delta y =0.2\lambda_{1}$, $\lambda_{1}=2\pi/k_{1}$. (a) $a=-1$, $b=0$; (b) $a=-1/\sqrt{2}$, $b=1/\sqrt{2}$; (c) $a=-0.2$, $b=0.98$; (d) $a=0.2$, $b=0.98$; (e) $a=1/\sqrt{2}$, $b=1/\sqrt{2}$; (f) $a=1$, $b=0$; (a1) is the same as (a) in 3D representation; (e1) is the same as (e) in 3D representation;}
\label{tape_3D}
\end{figure*}

In this subsection, we focus on studies of position distributions assuming that the cavity modes are in given reference state. Two schemes of joint measurements have been proposed up to now, based on a quadrature measurement of standing wave fields as well as on a measurement of the phase states \cite{Frey,Storey1992}. We start by considering the case of the quadrature measurement, the other case will be considered in the section \ref{Raman}. To realize this we consider implementation of a quadrature measurement on the field and use the following expression for the field quadrature state 
\begin{equation}\label{chiVec}
|\chi_{\theta}\rangle=(2\pi)^{-1/4}
e^{-(a^{\dag}e^{i\theta}-\chi_{\theta})^{2}/2+\chi_{\theta}^{2}/4}|vac\rangle.
\end{equation}
Here, parameter $\theta$ is an angle characterizing the one-mode field quadrature in Wigner plane, $\chi_{\theta}$ is the corresponding eigenvalue and $|vac\rangle$ denotes the vacuum state with zero photon in the cavity. Here we are interested in calculation of the probability of finding an atom at position $(x, y)$ provided that a measurement of the two field modes with angles $\theta_{1}$ and $\theta_{2}$ is performed. Using the reference state as $|\chi_{\theta_{1}}\rangle_{1}|\chi_{\theta_{2}}\rangle_{2}$ where quadrature state $|\chi_{\theta_{i}}\rangle_{i}$ corresponds to the operator $a_{i}$, we obtain the joint probability as
\begin{eqnarray}\label{VisPosDis}
W(\chi_{\theta_{1}},\chi_{\theta_{2}},x,y)=\sum_{i=1,2}|\prescript{}{1}{\langle}\chi_{\theta_{1}}|\prescript{}{2}{\langle}\chi_{\theta_{2}}|\langle i|\langle x,y|\Psi(t)\rangle|^{2}\nonumber\\
=\sum_{i=1,2}\Bigg|\sum_{n,m=0}^{\infty}\Phi_{n,m}^{(i)}(x,y,t)\prescript{}{1}{\langle}\chi_{\theta_{1}}|n\rangle_{1}\prescript{}{2}{\langle}\chi_{\theta_{2}}|m\rangle_{2}\Bigg|^{2}.
\end{eqnarray}
We obtain the general expression for the factor $\langle\chi_{\theta}|n\rangle$ using the formula (\ref{chiVec}). This matrix element can be calculated as follows
\begin{equation}
\langle \chi_{\theta}|n\rangle=\frac{1}{\pi}\int\langle \chi_{\theta}|\alpha\rangle\langle \alpha|n\rangle d^{2}\alpha, 
\end{equation}
where
\begin{equation}
\langle\chi_{\theta}|\alpha\rangle=\frac{e^{-|\alpha|^{2}/2}}{(2\pi)^{1/4}}\cdot e^{-(\alpha e^{-\mathbf{i}\theta}-\chi_{\theta})^{2}/2+\chi_{\theta}/4}
\end{equation}
and
\begin{equation}
\langle\alpha|n\rangle=e^{-|\alpha|^{2}/2}\cdot\frac{(\alpha^{\ast})^{n}}{\sqrt{n!}}.
\end{equation}
Similarly, the factors corresponding to the states $|\chi_{\theta_{1}}\rangle_{1}$ and $|\chi_{\theta_{2}}\rangle_{2}$ can be calculated. 

In the following, we show how various initial atomic states in the form $a|1\rangle+b|2\rangle$, $|a|^{2}+|b|^{2}=1$, with definite quadrature measurement can change the joint probability. Some typical results are depicted in Fig.\ref{rotation} as 2D distributions for the various superposition states $a|1\rangle+b|2\rangle$ as well as for the joint probabilities in 3D representation. The examples are taken for the measurement angles $\theta_{1}=\theta_{2}=0$, $\alpha_{1}=\alpha_{2}=2$, $\chi_{\theta_{1}}=\chi_{\theta_{2}}=4$ and $\Delta x=\Delta y =0.2\lambda_{1}$ ($\lambda_{1}=2\pi/k_{1}$). Considering 2D position distributions at the fixed parameters $\theta_{1}=\theta_{2}$, $\alpha_{1}=\alpha_{2}$, we conclude that the atomic distributions are turned in $x-y$ plane around the center $x=y=0$ in dependence on the coefficients $a$ and $b$. The figures showing these features are presented in the Fig.\ref{rotation}(a-f). 

As our analysis shows, the orientation of the distribution changes with the value of the fraction $a/b$ when it is varying from $-\infty$ to $\infty$ and any given orientation corresponds a particular value of $a/b$. In general, the fraction $a/b$ is a complex number and we can represent it as $(a/b)_{real}\cdot e^{\mathbf{i}\varphi}$, where $\varphi\in(-\pi/2, \pi/2)$ and $(a/b)_{real}\in(-\infty, \infty)$. It turns out that the additional factor $e^{i\varphi}$ does not change the orientation of the spatial distribution but changes the level of its stretching out in the direction of orientation. Thus, the distributions corresponding to $(a/b)_{real}$ and $(a/b)_{real}\cdot e^{\mathbf{i}\varphi}$ are qualitatively the same. This fact provides a possibility to calculate the $(a/b)_{real}$ fraction from a given final atomic position distribution.

Thus, we demonstrate on Fig.\ref{rotation} that atomic superposition state can be qualitatively probed in two-dimensional patterns of deflected atoms. Indeed, the distribution corresponding to the superposition state $\frac{1}{\sqrt{2}}(|1\rangle+|2\rangle)$ (Fig.\ref{rotation}(e)) is turned on $\pi/4$ relatively to the position distribution of $\Lambda$ - atoms that are initially in the state $|1\rangle$ (Fig.\ref{rotation}(a,f)).

As our analysis shows, for higher values of the parameters $\alpha_{1}$ and $\alpha_{2}$, than those that have been used in our example, the form of distributions are essentially the same. The slight differences are related to additional relatively small structures that do not change the spatial orientation features of the distribution. Thus, obtained results are applicable also for more intensive electromagnetic fields. The choice of particular values $\theta_{1}=\theta_{2}=0$ of the parameters $\theta_{1}$ and $\theta_{2}$ are conditioned by the explicitness of effects described above. The above results are mainly valid also for the other values of measurement angles $\theta_{1}$ and $\theta_{2}$ although differ in details from the results presented in Figs.\ref{rotation}.       

\subsection{Position distribution in the far-diffraction zone}

For the completeness of our analysis, we added in this section calculations for the position distribution of atoms in the far-field diffraction zone. This analysis is important for an experimental verification of the obtained effects. It allows us to consider the obtained effects for realistic experimental conditions and hence defining the spatial resolution necessary for detection of these effects. 

To modify previous calculations we use the expression for the free Hamiltonian of atoms $H_{free}=\hat{p}^{2}/(2m)$ and rewrite the state vector of system at time $t+t_{0}$, where $t$ is the free propagation time, in the following form    
\begin{eqnarray}\label{farVector}
|\Psi(t+t_{0})\rangle=\exp\left[-\frac{\mathbf{i}}{\hbar}H_{free}t\right]|\Psi(t_{0})\rangle\nonumber\\
=\int d x d y\sum_{i=1,2}\sum_{n,m=0}^{\infty}\bar{\Phi}_{n,m}^{(i)}(x,y,t)|n\rangle_{1}|m\rangle_{2}|i\rangle|x,y\rangle,
\end{eqnarray}
where
\begin{eqnarray}\label{farPhi}
\bar{\Phi}_{n,m}^{(i)}(x,y,t)=-\frac{\mathbf{i}M}{2\pi\hbar t}\int dx' dy' \Phi_{n,m}^{(i)}(x',y',t_{0})\nonumber\\
\times\exp\left[\frac{\mathbf{i}M}{2\hbar t}\left((x-x')^{2}+(y-y')^{2}\right)\right].
\end{eqnarray}

\begin{figure}[]
\begin{math}
\begin{array}{cc}
   \includegraphics[width=4.2cm]{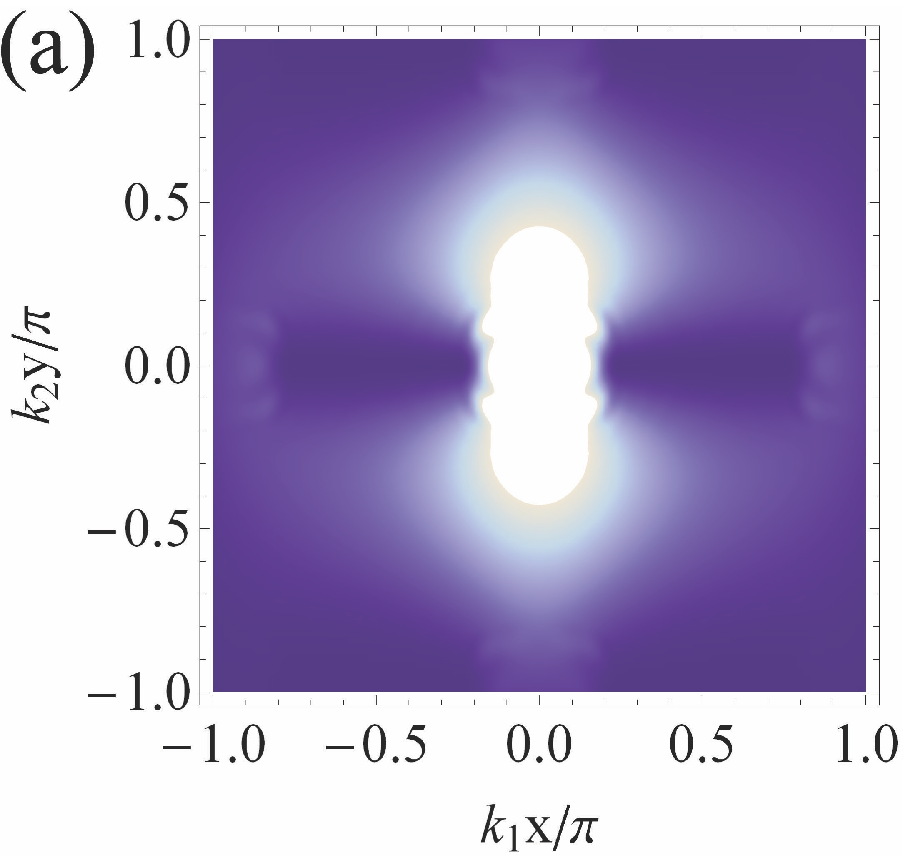} &  \includegraphics[width=4.2cm]{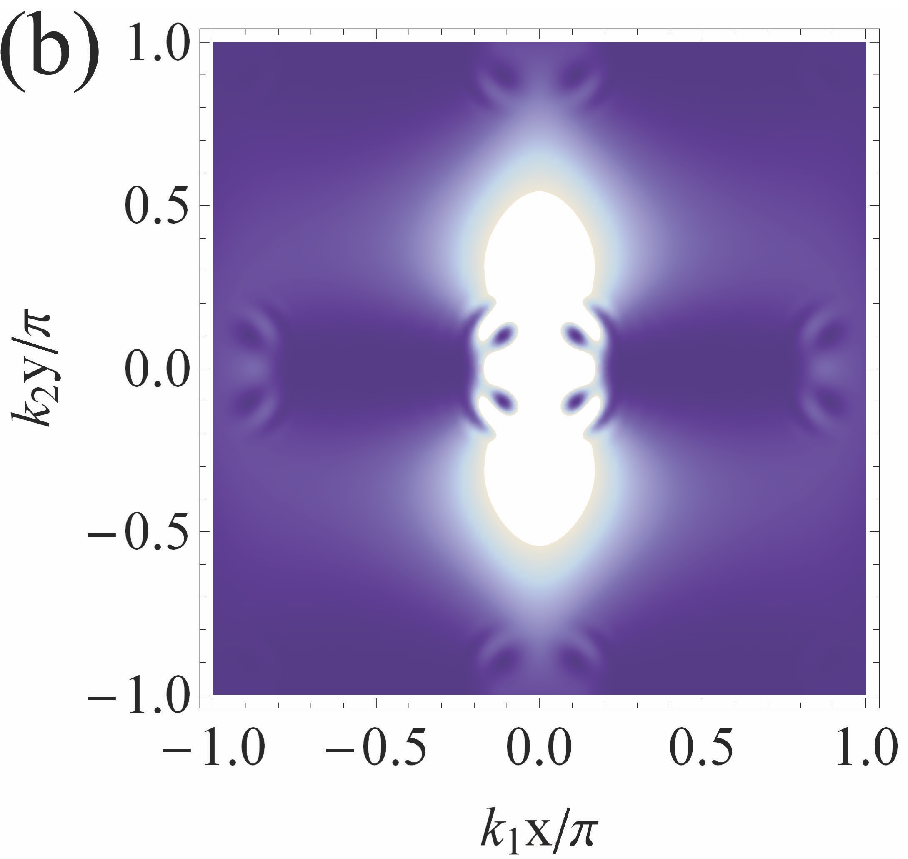}\\
\end{array}
\end{math}
\caption{\label{Far}The position distribution $W(x,y)$ for $\Delta x=\Delta y =0.2\lambda_{1}$, $\lambda_{1}=2\pi/k_{1}$ and initial internal state of the atom $a=1$ and $b=0$. Brighter colors indicate higher values of distribution. Distribution (a) at the exit of cavity; (b) at distance $L=vt=50cm$ from the exit of cavity.}
\end{figure}

From (\ref{farVector}) it is clear that for calculations of the position distribution at far-field diffraction zone we can use the expression  (\ref{VisPosDis}) just by replacing $\Phi_{n,m}^{(i)}(x,y,t)$ to $\bar{\Phi}_{n,m}^{(i)}(x,y,t)$. The results of numerical calculations of the distributions at near- and far-field regions are depicted in Figs.\ref{Far} showing a slight difference between them. The distribution in Fig.\ref{Far}(b) corresponds to distance $L=vt=50cm$ from the exit of cavity, where $v$ is the velocity of atomic mass center in $z$ direction and for its value we have chosen $100m/s$.  

\subsection{Momentum distributions}

The other approach analysing the atomic deflection process concerns to momentum distributions of atoms. In this section, we shortly discuss the momentum distribution in the deflected patterns of the Gaussian atomic wavepacket assuming the width of the wavepacket
to be much smaller than the wavelength of the modes. We calculate the probability of finding an atom with the transverse momentum $(p_{x}, p_{y})$ provided that a measurement of the two field modes with angles $\theta_{1}$ and $\theta_{2}$ is performed. This procedure is similar to the calculation of conditional position distribution made above. Thus, momentum distribution is written as
\begin{eqnarray}\label{VisMomDis}
P(\chi_{\theta_{1}},\chi_{\theta_{2}},p_{x},p_{y})=\sum_{i=1,2}|\prescript{}{1}{\langle}\chi_{\theta_{1}}|\prescript{}{2}{\langle}\chi_{\theta_{2}}|\langle i|\langle p_{x},p_{y}|\Psi(t)\rangle|^{2}\nonumber\\
=\sum_{i=1,2}\Bigg|\sum_{n,m=0}^{\infty}\tilde{\Phi}_{n,m}^{(i)}(p_{x},p_{y},t)\prescript{}{1}{\langle}\chi_{\theta_{1}}|n\rangle_{1}\prescript{}{2}{\langle}\chi_{\theta_{2}}|m\rangle_{2}\Bigg|^{2},
\end{eqnarray}
where the amplitude in the momentum space is calculated by the Fourier transformation over spatial variables
\begin{eqnarray}
\tilde{\Phi}_{n,m}^{(i)}(p_{x},p_{y},t)=\frac{1}{2\pi}\int\int dxdy \Phi_{n,m}^{(i)}(x,y,t)\nonumber\\
\times\exp[-\frac{\mathbf{i}}{\hbar}(p_{x}x+p_{y}y)]. 
\end{eqnarray}

\begin{figure}[]
\begin{math}
\begin{array}{cc}
   \includegraphics[width=4.2cm]{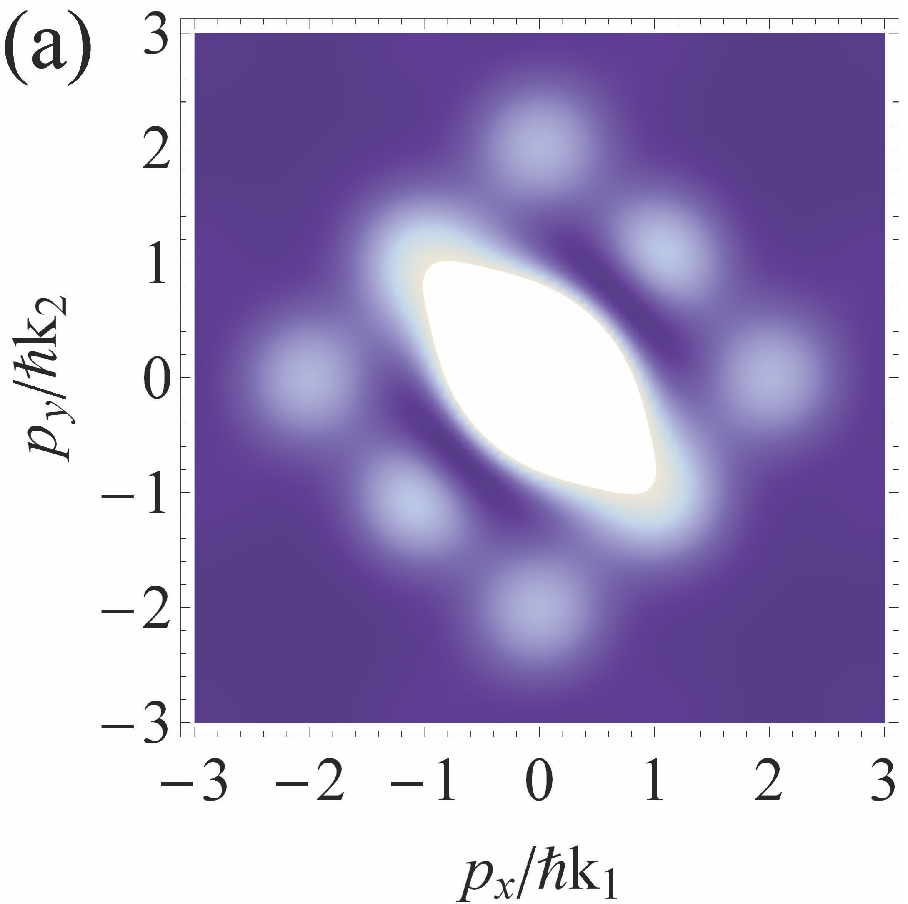} &  \includegraphics[width=4.2cm]{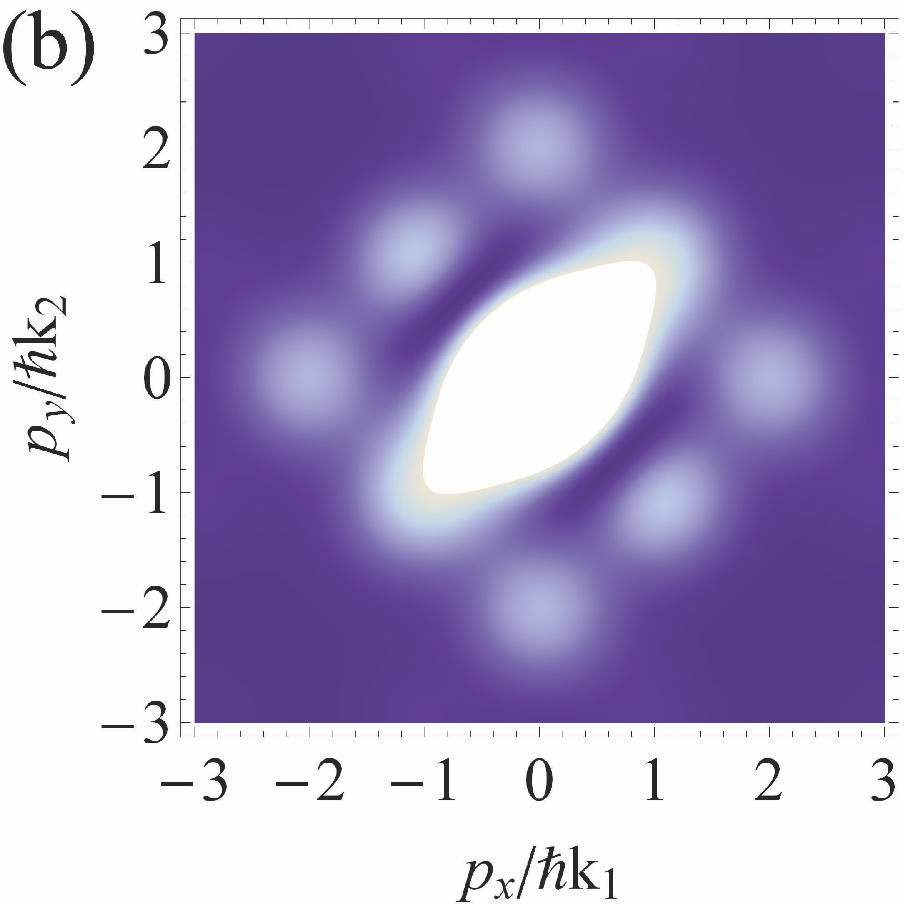}\\
\end{array}
\end{math}
\caption{\label{MomGraph}The momentum distribution $P(p_{x},p_{y})$ for $\Delta x=\Delta y =0.2\lambda_{1}$, $\lambda_{1}=2\pi/k_{1}$. Brighter colors indicate higher values of distribution. The initial internal states of the atom are:  (a) $a=-1/\sqrt{2}$ and $b=1/\sqrt{2}$, (b) $a=1/\sqrt{2}$ and $b=1/\sqrt{2}$. We use the dimensionless momenta scaled in units of $\hbar k$.}
\end{figure}

We illustrate the momentum distributions for narrow
initial atomic wave packets which obviously corresponds to a wider distribution in the momentum space.
As shows analysis, in contrast to the position distributions the dependence of momentum distributions on initial internal states is not so evident, but in some cases it is still possible to find visible relations between distribution and initial internal state. The momentum distributions for two cases of initial atomic superposition states with the coefficients $a=-1/\sqrt{2}$, $b=1/\sqrt{2}$ and $a=1/\sqrt{2}$, $b=1/\sqrt{2}$ are shown in Fig.\ref{MomGraph}. Comparing Fig.\ref{MomGraph}(a) and Fig.\ref{MomGraph}(b) we realize  that  ranges of localizations of the momentum for these cases are stretched in perpendicular directions, i.e. the distribution corresponding to the superposition state $\frac{1}{\sqrt{2}}(|1\rangle+|2\rangle)$  is turned on $\pi/2$ relatively to the momentum distribution of $\Lambda$ - atoms that are initially in the state $\frac{1}{\sqrt{2}}(|1\rangle-|2\rangle)$.  Comparing the results of momentum distributions with  corresponding position distributions in Fig.\ref{rotation}(b) and Fig.\ref{rotation}(e) we conclude that they are in accordance with uncertainty relations. 

\section{\label{Raman}Evidence of Raman resonance in atomic deflection}

In this section we illustrate the principal differences between the cases of interaction in Raman resonance and ordinary off-resonance from the point of view of an atomic beam deflection by two crossed standing wave. In this way, we derive the amplitudes $\Phi_{n,m}^{(i)}(x,y,t)$ in (\ref{FinalS}) for interaction Hamiltonian (\ref{OutRaman}). For atomic initial internal state $a|1\rangle+b|2\rangle$ the amplitudes are written as 
\begin{subequations}
\label{outRPhis}
\begin{eqnarray}
& \Phi_{n,m}^{(1)}(x,y,t)=f(x,y)C_{n,m}\exp\left[-\mathbf{i}\frac{g_{1}^{2}t}{\Delta_{1}}n\right]\cdot a,\\
& \Phi_{n,m}^{(2)}(x,y,t)=f(x,y)C_{n,m}\exp\left[-\mathbf{i}\frac{g_{2}^{2}t}{\Delta_{2}}m\right]\cdot b.
\end{eqnarray}
\end{subequations}
Considering different regimes of interaction of $\Lambda$ - atom by using the spatial amplitude (\ref{Phis}), (\ref{ABs}) and (\ref{outRPhis}) we demonstrate that the deflection patterns are essentially different for the case of one-photon and two-photon interactions. The most convenient approach to realize our goal is the investigation of conditional position distribution of atoms provided that the two mode field is in a given phase state   
\begin{equation}
|\Psi_{R}\rangle=|\varphi_{1}\rangle_{1}|\varphi_{2}\rangle_{2},
\end{equation}
where
\begin{equation}
|\varphi\rangle_{i}=\frac{1}{\sqrt{2\pi}}\sum_{n}e^{-\mathbf{i}\varphi n}|n\rangle_{i}.
\end{equation}
The discussing distribution is
\begin{equation}\label{PosDisRef}
W(x,y)=\sum_{i=1,2}|\langle x,y|\langle i|\langle\Psi_{R}|\Psi(t)\rangle|^{2}.
\end{equation}

Using the formulas (\ref{FinalS}) and (\ref{outRPhis}-\ref{PosDisRef}) we obtain the folloing expression for $W(x,y)$
\begin{equation}\label{disVarphi}
W(x,y)=\frac{1}{(2\pi)^{2}}\sum_{i=1,2}\bigg|\sum_{n,m}\Phi_{n,m}^{(i)}(x,y,t)\cdot e^{-\mathbf{i}(n\varphi_{1}+m\varphi_{2})}\bigg|^{2}.
\end{equation}
Obviously, distributions (\ref{outRPhis}) and (\ref{disVarphi}) crucially depend on phases $\varphi_{1}$ and $\varphi_{2}$ which concretize the ranges of atom–waves interactions. We choose $\varphi=0$, which gives the best spatial localization of the scattered atoms. Thus, the probability $W(x,y,\varphi=0)=W(x,y)$ describes the selected scattering events of only those atoms that have passed the nodes of the field with the spatial mode function $\sin(k_{1}x)$, $\sin(k_{2}y)$. As it was shown earlier, such a joint measurement procedure plays a role of a spatial filter. Experimentally, this procedure can be realized by a mechanical slit or mask placed in front of a node of the electromagnetic field. For detailed discussion of this point, see \cite{Frey}.
 
\begin{figure}[]
\begin{math}
\begin{array}{cc}
   \includegraphics[width=4.2cm]{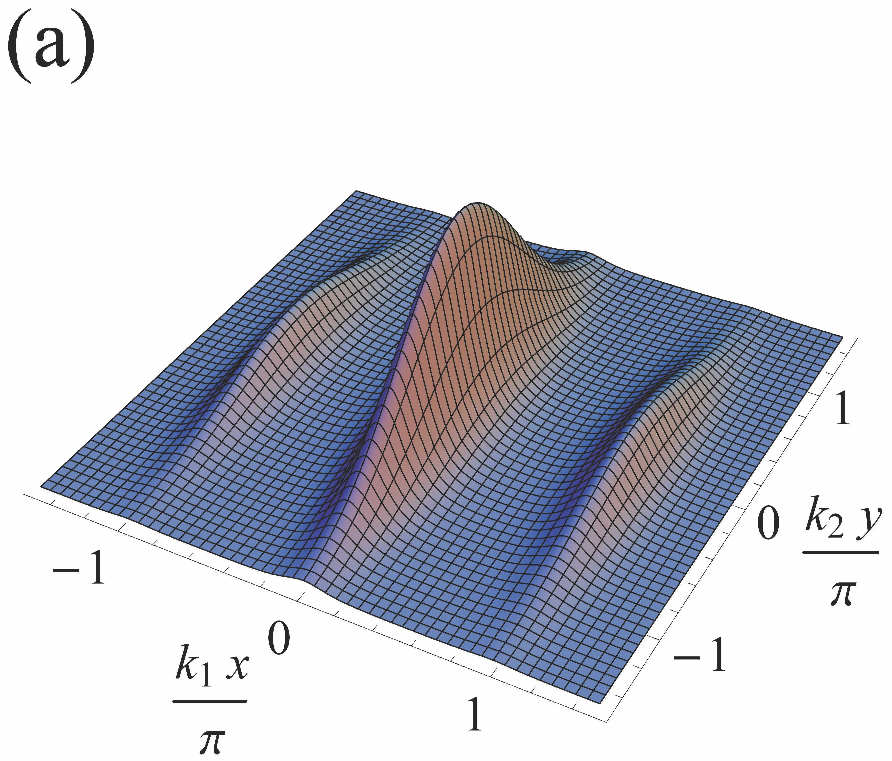} & \includegraphics[width=4.2cm]{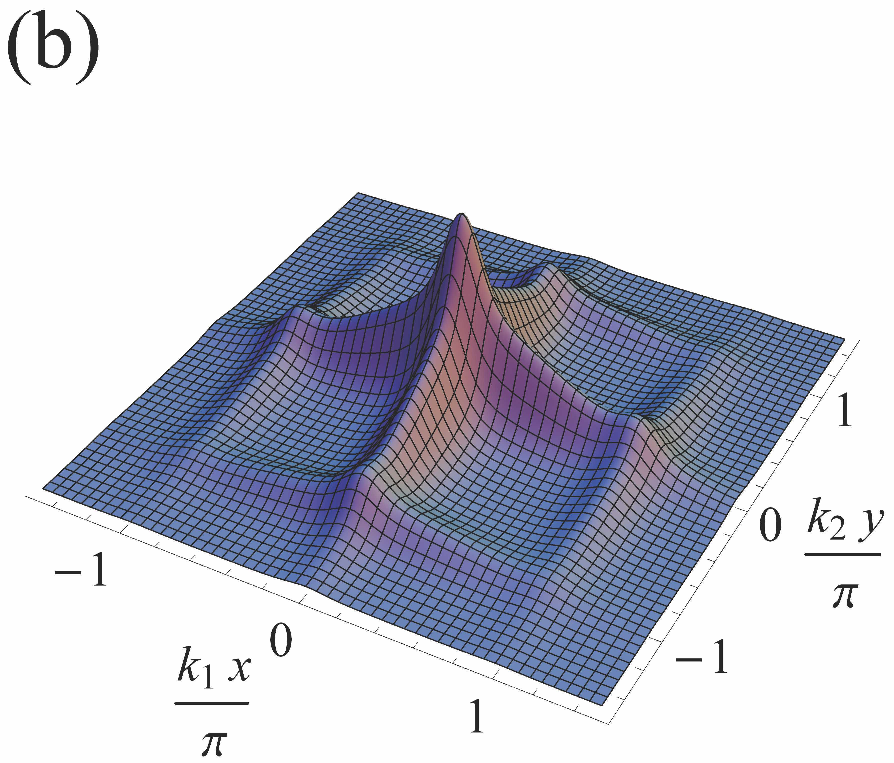}\\
\end{array}
\end{math}
\caption{\label{OutRamGraph}The position distribution function $W(x,y)$ for the case of $\Delta_{1}\neq\Delta_{2}$, $\Delta x=\Delta y =0.3\lambda_{1}$, $\lambda_{1}=2\pi/k_{1}$. Initial state of the atom is described by (a) $a=1$ and $b=0$; (b) $a=1/\sqrt{2}$ and $b=1/\sqrt{2}$.}
\end{figure}

\begin{figure}[]
\begin{math}
\begin{array}{cc}
   \includegraphics[width=4.2cm]{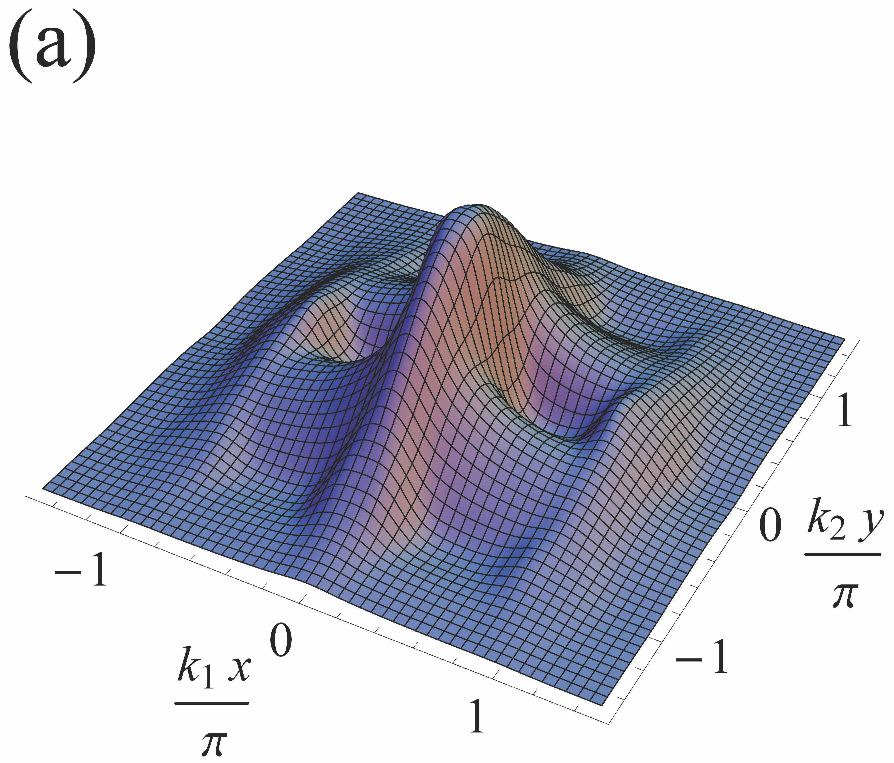} & \includegraphics[width=4.2cm]{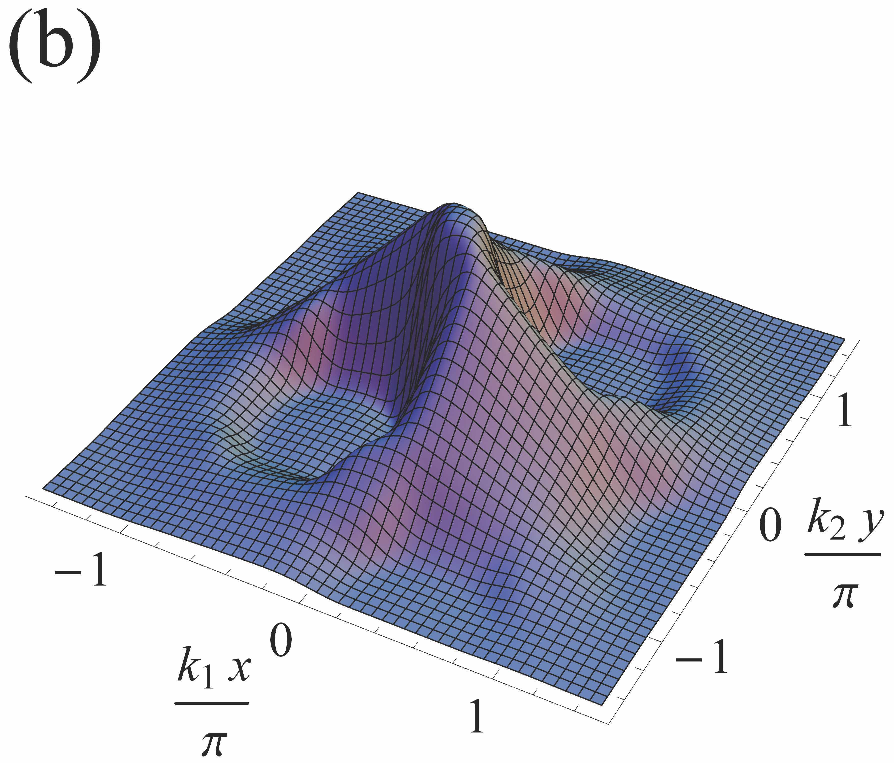}\\
\end{array}
\end{math}
\caption{\label{RamanGraph}The position distribution function $W(x,y)$ for the case of $\Delta_{1}=\Delta_{2}$, $\Delta x=\Delta y =0.3\lambda_{1}$, $\lambda_{1}=2\pi/k_{1}$. Initial state of the atom is described by (a) $a=1$ and $b=0$; (b) $a=1/\sqrt{2}$ and $b=1/\sqrt{2}$.}
\end{figure}

The results of concrete calculations based on Eq. (\ref{disVarphi}) are depicted on the Fig.\ref{OutRamGraph} and Fig.\ref{RamanGraph} for the different regimes of interaction. In Fig.\ref{RamanGraph} the position distribution functions $W(x,y)$ are depicted in the case of two-photon resonant interaction and for two different choices of the initial coherence of the internal atomic state. As we can see, these results are similar to those in the Fig.\ref{rotation}(a1,e1) obtained for the quadrature-measurement scheme. The case of different detunings leading to one-photon atomic transitions is depicted in Fig.\ref{OutRamGraph}. Comparing two types of the graphics allows us to make clear the peculiarities of coherent Raman processes in comparison with corresponding one-photon processes from the point of view of atomic optics.
  
Let us consider probability distributions depicted in Fig.\ref{OutRamGraph}(a) and Fig.\ref{RamanGraph}(a) that correspond to the situation when atoms enter  cavity  being  initially in the lower state $|1\rangle$. As we see, the position distribution in Fig.\ref{OutRamGraph}(a) shows high spatial localization of deflected atoms  in $x$ direction, while in $y$ direction their behaviour have remained unchanged, i.e. determined by the initial Gaussian distribution. In other words, the impact of the initial field state is the one-dimensional distribution of the atoms. It could be anticipated, because in this case the atom and the field interact through only one channel, (it is the interaction between first mode of the field and $|1\rangle\leftrightarrow|3\rangle$ transition), as the atom due to off-resonant nature of the interaction always remains in the state $|1\rangle$. In contrast to that, in the case of two-photon resonance the two-photon $|1\rangle\leftrightarrow|2\rangle$ transition leads to activation of the interaction between the second mode of the field and the $|2\rangle\leftrightarrow|3\rangle$ transition. As a result, the structure of the position distribution in Fig.\ref{RamanGraph}(a) has acquired some features indicating this fact:  The distribution showing the localization near some points in both $x$ and $y$ directions is described by two-dimensional patterns. Particularly, it is expressed by emergence of four additional “walls” forming a structure resembling two closed cycles (see, Fig.\ref{RamanGraph}(a)).

If the atom enters the cavity in a superposition of two lower $|1\rangle$ and $|2\rangle$ states, the impact of the field on atomic position distributions contains two-dimensional localization patterns even in case of the absence of Raman transitions (see Fig.\ref{OutRamGraph}(b) and Fig.\ref{RamanGraph}(b)). The difference between the distributions of considered two regimes becomes obvious when we analyze the graphics in the following manner. If Raman resonance is violated the position distribution (see, Fig.\ref{OutRamGraph}(b)) has described by two perpendicular planes of symmetry passing through $x$ and $y$ axes. It expresses the fact that the impact of interaction between the first mode and $|1\rangle\leftrightarrow|3\rangle$ transition and the impact of interaction between the second mode and $|2\rangle\leftrightarrow|3\rangle$ transition are independent. They depend only on populations on the corresponding atomic levels, which in this regime, at the beginning of interaction are equal to $|a|^{2}$ and $|b|^{2}$ respectively, (see Eq.(1)). In the two-photon resonance regime $|1\rangle\leftrightarrow|2\rangle$ transition destroys this independence and hence, leads to breaking of the symmetry. This fact is evidently reflected in Fig.\ref{RamanGraph}(b) where $x$ and $y$ are not axes of symmetry for the distribution.

\section{\label{Sum}Summary}

By calculating the concrete conditional spatial distributions of $\Lambda$-structured atoms after passing two crossed standing light waves we have shown that two-dimensional patterns of deflected atoms contain important information concerning the low-level atomic superposition states as well as reflect the efficiency of the two-photon resonant Raman process. Thus, we have developed an approach for testing and visualization of superposition states as well as for probing the Raman resonance in a new not spectroscopic manner. In more details, we have studied how various initial atomic superposition states $a|1\rangle+b|2\rangle$ in the Gaussian atomic beam are visualized in both 2D and 3D distributions of the joint probabilities. In this way, we have demonstrated that at the fixed values of the parameters $\theta_{1}=\theta_{2}$, $\alpha_{1}=\alpha_{2}$ the atomic spatial patterns are turned in $x-y$ plane around the center $x=y=0$ on an angle depending on the values of the coefficients $a$ and $b$ describing the weights of the
atomic lower states probabilities in the initial superposition state. This analyses has been done for both near- and far-field diffraction zones as well as in the momentum space. Considering two regimes of the interaction of  $\Lambda$-type atoms with two-mode field we have clearly demonstrated on the spatial patterns the peculiarities of coherent Raman processes in comparison with corresponding one-photon processes from the point of view of atomic optics.

\begin{acknowledgments}
G. Yu. K. acknowledges helpful discussions with J. Evers, Ch. H. Keitel, M. Macovei and M. Zubairy. The authors acknowledge collaboration grant between the Armenian and Hungarian Academies of Sciences. G.P.D. acknowledges support of the Research Fund of the Hungarian Academy of Sciences (OTKA) under contracts K 68240, NN 78112, as well as of the ELI-09-1-2010-0010 grant.
\end{acknowledgments}


\bibliographystyle{apsrev4-1}
\bibliography{abovyan}

\end{document}